# MANIFESTATION OF STRONG QUADRUPOLE LIGHT-MOLECULE INTERACTION IN THE SER AND SEHR SPECTRA OF PYRAZINE AND PHENAZINE

A.M. Polubotko

A.F. Ioffe Physico-Technical Institute Russian Academy of Sciences, Politechnicheskaya 26

194021 Saint Petersburg RUSSIA, Fax: (812) 297-10-17, E-mail: alex.marina@mail.ioffe.ru

## ABSTRACT

It is demonstrated that explanation of appearance of strong forbidden lines in the SEHR spectra of pyrazine and phenazine, caused by totally symmetric vibrations and also other details of their SER and SEHR spectra can be made on the base of the dipole-quadrupole theory. The main point of this theory is conception of a strong quadrupole light molecule interaction, arising in surface fields strongly varying in space near a rough metal surface. Existence of the pointed lines is a good corroboration of existence of the strong quadrupole light-molecule interaction.

In accordance with main principles of theoretical physics, interaction of light with molecules is determined by light molecule interaction Hamiltonian, which has the following form

$$\widehat{H}_{e-r} = -\sum_i \frac{ie\hbar}{mc} \overline{A}_i \nabla_i ,  \qquad (1)$$

where summation is made over all $i$ electrons. $\overline{A}_i$ is a vector potential at the place of the $i$ electron. All other designations are conventional. For small objects like molecules, the expression (1) can be transformed and neglecting by the magneto-dipole interaction the final expression for (1) has the form

$$\hat{H}_{e-r} = |\overline{E}| \frac{(\overline{e}^* \overline{f}^*)e^{i\omega t} + (\overline{e}\,\overline{f})e^{-i\omega t}}{2}, \qquad (2)$$

where $|\overline{E}|$ is the amplitude of the electric field in the gravity centre of the molecule,

$$f_\alpha = d_\alpha + \frac{1}{2E_\alpha} \sum_\beta \frac{\partial E_\alpha}{\partial x_\beta} Q_{\alpha\beta} \qquad (3)$$

is an $\alpha$ component of a generalized coefficient of interaction of light with electrons (molecule), $\overline{e}$ is a polarization vector, $d_\alpha$ and $Q_{\alpha\beta}$ are the $\alpha$ component of the dipole moment vector and the $\alpha\beta$ component of the quadrupole moments tensor of electrons. As it is demonstrated in [1, 2] both terms in (3) strongly increase near prominent places with a large curvature, which exist near a rough metal surface. This increase is associated with singular behavior of a diffracted electromagnetic field near such model of roughness as a cone or with a rod effect.

$$E_{r'} \sim |\overline{E}_{inc}|_{vol} C_0 \left(\frac{l_1}{r'}\right)^\beta. \qquad (4)$$

Here $|\overline{E}_{inc}|_{vol}$ is an amplitude of the incident electric field in a free space, $C_0 \sim 1$ is a numerical coefficient, $l_1$ is a characteristic size of the cone (the size of the base or the height), $r'$ is the length of the radius vector from the top of the cone, $0 < \beta < 1$. As it can be seen from (4) there is a very large increase of the component of the electric field, which is perpendicular to the surface, while the tangential components are small because of screening of the field by conductivity electrons.

Taking into account (3) the expression (2) can be presented as

$$\widehat{H}_{e-r} = \widehat{H}_d + \widehat{H}_Q, \tag{5}$$

where $\widehat{H}_d$ contains only the dipole terms, while $\widehat{H}_Q$ contains the quadrupole terms. The enhancement of the dipole and quadrupole light-molecule interactions in (2, 5) is associated with the increase of the dipole terms and quadrupole terms with $\alpha = \beta$. For the model roughness as a finite cone the enhancement coefficients $G_{H_d}$ and $G_{H_Q}$ near the top of the cone, which are the relations of the corresponding terms of the Hamiltonian (5) to the Hamiltonian in a free space can be estimated as

$$G_{H_d} \sim C_0 \left(\frac{l_1}{r'}\right)^\beta, \tag{6}$$

$$G_{H_Q} \sim C_0 \beta \left(\frac{B_{\alpha\alpha}}{2}\right)\left(\frac{l_1}{r'}\right)^\beta \left(\frac{a}{r'}\right), \tag{7}$$

where $a$ is a size of the molecule. Peculiarity of $G_{H_Q}$ is appearance of the coefficients

$$B_{\alpha\alpha} = \frac{\overline{\langle m|Q_{\alpha\alpha}|n\rangle}}{\overline{\langle m|d_\alpha|n\rangle}} \gg 1, \tag{8}$$

which are the relations of some mean values of the matrix elements of the quadrupole moments $Q_{\alpha\alpha}$ and the dipole moments $d_\alpha$ and arise due to peculiarities of quadrupole transitions via the $Q_{\alpha\alpha}$ quadrupole moments, which are the values with a constant sign. As it was demonstrated in [1, 2] these coefficients are large and their estimations for the molecules like pyridine, benzene and pyrazine give the value $\sim 2 \times 10^2$. This value depends on the size of the molecule and is larger for larger molecules. As it can be seen from (6, 7) there is a strong enhancement of both interactions in the points, which are close to the top of the cone ( $r' \sim 0$ ) and a strong decrease at larger distances. For example for reasonable values of the parameters

$C_0 \sim 1$, $l_1 \sim 10\,nm$, $r' \sim 1\,nm$, $\beta \sim 1$ and $B_{\alpha\alpha} \sim 2\times 10^2$ the enhancement of the dipole interaction $\sim 10$, while the enhancement of the quadrupole interaction $\sim 10^2$. However for some limited situation and the values of the parameters $C_0 \sim 1$, $B_{\alpha\alpha} \sim 2\times 10^2$, $r' \sim 0.1\,nm$, $\beta \sim 1$, $l_1 \sim 100\,nm$ that corresponds to the placement of the molecule on the top of the cone (tip or spike) the enhancement coefficient $G_{H_d} \sim 10^3$, while $G_{H_Q} \sim 10^5$. Since SERS and SEHRS are the processes of the second and the third orders respectively the estimation of the enhancement coefficients for the dipole and quadrupole enhancement mechanisms are

$$G_d \sim C_0^{2n} \left(\frac{l_1}{r'}\right)^{2n\beta}, \tag{9}$$

$$G_Q \sim C_0^{2n} \beta^{2n} \left(\frac{B_{\alpha\alpha}}{2}\right)^{2n} \left(\frac{l_1}{r'}\right)^{2n\beta} \left(\frac{a}{r'}\right)^{2n} \tag{10}$$

where $n = 2$ and $n = 3$ for SERS and SEHRS respectively. Their numerical estimation for the limited case gives $10^{12}$ and $10^{20}$ for SERS and $10^{18}$ and $10^{30}$ for SEHRS respectively. One should note that these estimations are very crude. However they can give an idea about mechanism and huge enhancement of these processes. In addition we should point out that analogous estimation was done for the crystal violet molecule in SERS in the single molecule regime [3] that gives the enhancement coefficient in SERS $\sim 1.6\times 10^{16}$. This estimation is huge, but significantly less, than the above one. The estimation $10^{20}$ can be raised too high, because of the some limited case with a very difficult realization.

On Figures 1 and 2 we present the distance dependence of the relative enhancement in SERS and SEHRS normalized to the maximum enhancement ($G_Q/G_{Q,\max}$ and $G_d/G_{d,\max}$) for the pure dipole and pure quadrupole scattering mechanisms for the cone type roughness. One on the x axis corresponds to a half of the height of adsorbed molecule, or the coordinate of the first monolayer. 3 corresponds to the coordinate of the second monolayer when the molecules are oriented in the same manner.

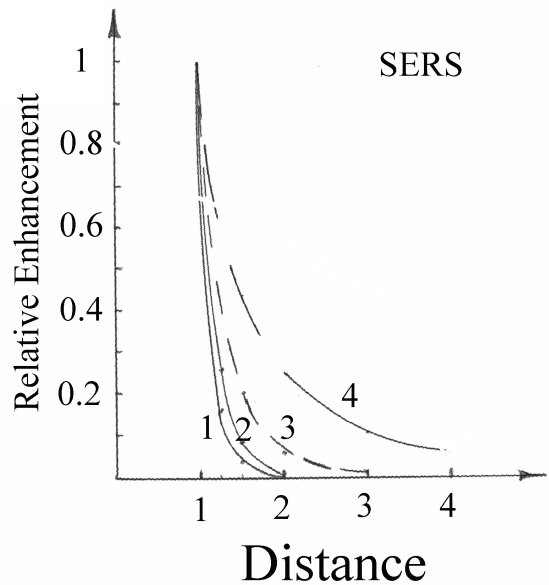

Figure 1 The distance dependence of the relative enhancements $G_Q / G_{Q,\max}$ and $G_d / G_{d,\max}$ for SERS for the pure dipole and quadrupole scattering mechanisms. Curves 1,2 –the dependences for the quadrupole mechanism for $\tau = 1$ and $\tau = 1/2$ respectively, curves 3,4 –the dependences for the dipole mechanism for $\tau = 1$ and $\tau = 1/2$.

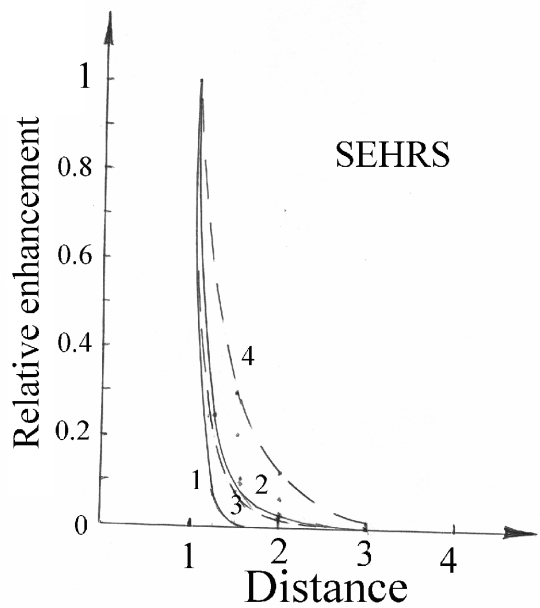

Figure 2 The distance dependence of the relative enhancement $G_Q / G_{Q,\max}$ and $G_d / G_{d,\max}$ for SEHRS for the pure dipole and quadrupole scattering mechanisms. Curves 1,2 –the dependences for the quadrupole mechanism for $\tau = 1$ and $\tau = 1/2$ respectively, curves 3,4 –the dependences for the dipole mechanism for $\tau = 1$ and $\tau = 1/2$.

It can be seen that the enhancement in both processes decreases very strongly with the distance from the top of the cone. In fact the maximum enhancement arises in the first layer, while in the second layer it decreases on orders of magnitude. We pointed out on this behavior earlier [4,5] and it is in a good qualitative agreement with the experimental distance dependence for SERS obtained in [6] (see Figure 3).

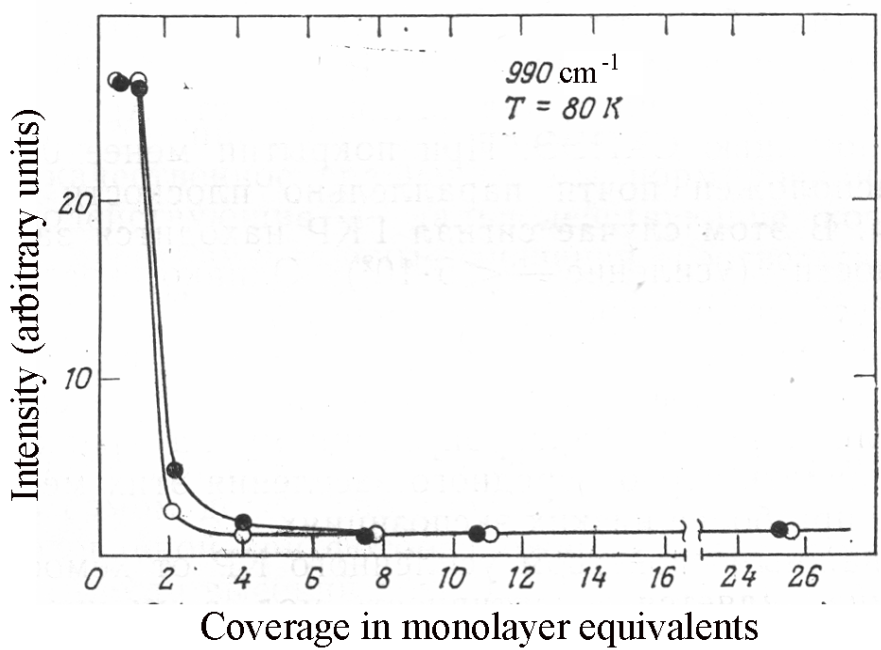

Figure 3 Experimental distance dependence for the enhancement

In addition this decrease is significantly stronger for SERHS than for SERS, because SEHRS is the process of larger order. In fact we see the so called first layer effect, which is well known in SERS. It should be noted, that this effect gave rise to the so called charge transfer, or chemical mechanism, which is based on the fact of the largest enhancement in the first layer and was connected by authors with the direct contact of the molecules with the surface. Now it is seen, that this effect has a pure electrodynamical nature and is associated with very strong change of the electric field and its derivatives with the distance from the top of the cone in our model of the roughness. In more realistic models of the roughness, when we have a truncated cone and the top of the cone has a finite curvature, this change is less, however in any case the large decrease preserves.

In accordance with the above estimations one can introduce a conception of main and minor moments. The main dipole and quadrupole moments are those, which are responsible for the strong enhancement. This classification essentially depends on orientation of the molecules near rough surface. Further we shall consider the pyrazine and phenazine molecules. Let us designate the coordinate system, which is associated with the molecules as $(x,y,z)$ and the one with the surface as $(x',y',z')$. We consider that the plane of the molecules coincide with the $XZ$ plane and the $z$ axis passes through nitrogen atoms. The $z'$ axis is perpendicular to the surface. Usually the experiments are performed on molecules in solutions and the substrate is under some negative applied potential. The molecules in the first layer can bind with the surface via the nitrogen lone pair having the end on orientation. Other molecules are in the solution and can have an arbitrary orientation. The number of molecules, which are bound with the surface apparently is large, because of the binding, which prevents their movement and arbitrary orientation in the solution. The negative potential prevents the binding. Therefore the number of the bound molecules apparently depends on the potential. The number of arbitrary oriented molecules strongly depends on their concentration in the solution. Apparently their number situated directly at the surface decreases with the increasing of the negative potential, because the binding becomes less energetically favorable and more molecules at the surface become free. Because of existence of molecules which are oriented arbitrary with respect to the surface and hence to the enhanced $E_{z'}$ component of the electric field all the $d_\alpha$ and $Q_{\alpha\alpha}$ quadrupole moments are essential for the scattering and are the main ones. As it is well demonstrated in [1], the other $Q_{xy}, Q_{xz}, Q_{yz}$ quadrupole moments are not essential, because of their changeable sign and we name them as minor moments. In order to receive an idea about the role of the dipole and quadrupole interactions in SERS and SEHRS it is reasonable to consider the SER and SEHR spectra of some symmetrical molecules, that can give precise information about allowed and forbidden bands. The symmetry group of the pyrazine and phenazine molecules, considered here is $D_{2h}$ and all the dipole and quadrupole moments transform after irreducible representations of the symmetry group. Precise expression for the SER cross-section for these molecules, which can be obtained in the frame of adiabatic perturbation theory [7, 8] by the method of time dependent perturbation theory is contained in [1] and in Appendix and has the form.

$$d\sigma_{SERS_{s,surf}}\begin{pmatrix}St\\AnSt\end{pmatrix} = \frac{\omega_{inc}\omega_{scat}^3}{16\hbar^2\varepsilon_0^2\pi^2c^4}\frac{\left|\overline{E}_{inc}\right|^2_{surf}}{\left|\overline{E}_{inc}\right|^2_{vol}}\frac{\left|\overline{E}_{scat}\right|^2_{surf}}{\left|\overline{E}_{scat}\right|^2_{vol}} \times$$
$$\times \begin{pmatrix}(V_s+1)/2\\V_s/2\end{pmatrix}\left|\sum_{f_i,f_j}S_{s,f_i-f_j}\right|^2_{surf}dO$$
(11)

Here it is taken into account, that all the states are non degenerated in the $D_{2h}$ symmetry group and the vibrational states can be classified only by one index $s$. All designations are explained in Appendix. As it is seen from (11) the scattering contributions $S_{s,f_i-f_j}$ obey selection rules ([1] and (A-38))

$$\Gamma_s \in \Gamma_{f_i}\Gamma_{f_j}.$$
(12)

The sign $\Gamma$ designates the irreducible representations of the $s$ vibration and of the $f_i$ and $f_j$ moments. Further we shall designate the scattering contributions for SERS simply as $(f_i - f_j)$. In accordance with our estimations of the role of the dipole and quadrupole interactions in SERS and SEHRS (6,7,9,10) the above contributions can be classified after enhancement degree. The sequence below reflects the decrease in the degree of the enhancement for the SER scattering contributions.

1. the strongest, essential $(Q_{main} - Q_{main})$ scattering type;

2. essential $(d_{main} - Q_{main})$ and $(Q_{main} - d_{main})$ scattering types;

3. essential $(d_{main} - d_{main})$ scattering types.

The contributions of the first type must be strongly enhanced, while the contributions of the second and the third types are also enhanced, but apparently to a lesser extent. The following contributions will be enhanced significantly lesser in case of the vertical and horizontal orientation of the molecules. Here they are written out in accordance with the enhancement degree.

4. $(d_{minor} - Q_{main})$ and $(Q_{main} - d_{minor})$ scattering types;

5. $(Q_{minor} - Q_{main})$ and $(Q_{main} - Q_{minor})$ scattering types;

6. $(d_{minor} - d_{main})$ and $(d_{main} - d_{minor})$ scattering types;

7. $(Q_{minor} - d_{main})$ and $(d_{main} - Q_{minor})$ scattering types.

Here $d_{minor}$ are the dipole moments, which are oriented parallel to the tangential components of the electric field $E_{x'}$ and $E_{y'}$. The other $(d-d), (d-Q), (Q-d)$ and $(Q-Q)$ contributions, in which both the dipole and quadrupole moments are of the minor type, are apparently small. Therefore we name them as inessential parts in SERS. One should note, that the scattering contributions can differ one from another very strongly. Therefore the intensity of the Raman bands can be determined mainly by the strongest contributions, while all others contribute insignificantly.

Since the $Q_{main}$ moments transform after the unit irreducible representation, the most enhanced bands in SERS are those transforming after the unit irreducible representation $A_g$ too, in accordance with the above classification and the selection rules (12). They are caused mainly by the $(Q_{main} - Q_{main})$ scattering contributions. In addition there must be strongly enhanced lines, caused by vibrations, transforming after the irreducible representations $B_{1u}$ $B_{2u}$ and $B_{3u}$ associated with the $(Q_{main} - d_{main})$ types of the scattering, which are forbidden in usual Raman scattering for molecules with the $D_{2h}$ group. One should note that the contributions of the $(d_{main} - d_{main})$ scattering types contribute to the lines with the $A_g$ and $B_{1g}, B_{2g}, B_{3g}$ irreducible representations. Thus the lines with nearly all irreducible representations except $A_u$ can appear in the SER spectra of the considered molecules. The SER spectra and assignment of the lines in pyrazine were described in various experiments [9-18]. Since the assignment depends on orientation of the pyrazine molecule with respect to the coordinate system, the same lines are assigned to various irreducible representations in [9-18]. Besides the wavenumbers in these papers differ one from another because of various experimental conditions, which were used. However the $B_{1u}, B_{2u}, B_{3u}$ modes transfer one through another in various coordinate systems. The same can be said about the $B_{1g}, B_{2g}, B_{3g}$ modes.

After analysis of the data of the SER spectra of pyrazine, published in [9-18] we can come to conclusion that the most enhanced bands are those caused by vibrations transforming after the unit irreducible representation $A_g$. There are both the $B_{1u}, B_{2u}, B_{3u}$ lines, which are forbidden in usual Raman spectrum of pyrazine and the $B_{1g}, B_{2g}, B_{3g}$ lines. The lines caused by the vibrations with $A_u$ irreducible representation are absent. As a specific example let us consider the SER spectrum of pyrazine published in [17]. The SER spectrum of pyrazine is presented as an upper curve on Fig. 4.

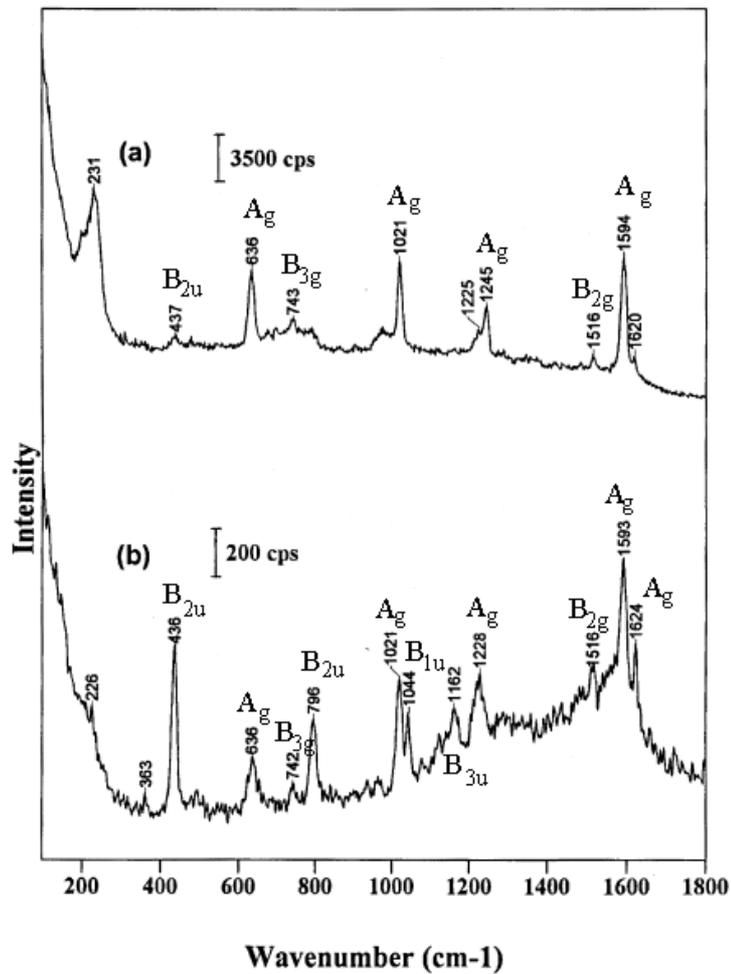

Figure 4 The SER **(a)** and SEHR **(b)** spectra of pyrazine.

In accordance with the results of [17] the most enhanced bands are 636, 1021, 1245 and 1594 $cm^{-1}$ caused by vibrations transforming after the unit irreducible representation $A_g$. Only one band of the $B_{2u}$ type at 437 $cm^{-1}$ caused by the ($Q_{main} - d_y$) type of the scattering of the horizontally and arbitrary oriented molecules manifests in this experiment. One should note that there are another lines of "u" type in another experiments [9-16, 18]. However the assignment of the same bands can differ one from another, because of the difference in orientation of the pyrazine molecule, which is used in these papers. In addition the bands at 743 and 1516 $cm^{-1}$ of the $B_{3g}$ and $B_{2g}$ symmetry types caused by $(d_z - d_y)$ and $(d_z - d_x)$ scattering contributions of arbitrary oriented molecules are also observed in [17]. The number of observed lines of the "g" types in experiments in [9-16, 18] is significantly more. Thus the pointed results are in a strict conformity with our theory.

Further we shall interpret results on the SER spectra of phenazine. One should note that the SER spectra were obtained in [19-21]. The distinctive peculiarity of these spectra is observation of overtones and combination bands. Besides the authors mention formation of the protonated forms of phenazine, like PH and $PH_2$ [21]. These forms manifest in the spectra by appearance of new lines. However one can convince that the lines of phenazine still remain, that points out its existence in the system and allows us to analyze its SER spectra. Separation of the lines, which refer to phenazine is possible due to their good assignment made in [22, 23]. Here we shall interpret the SER spectra of phenazine, obtained in [21] Fig. 5. In accordance with our theory the most enhanced bands are those which belong to the vibrations with $A_g$ symmetry. They are the bands with the wavenumbers 1567, 1406, 1282, 1170, 1021, 607, 413 $cm^{-1}$. Here we point out the bands, which arise at 0 V potential of the substrate (The curve (a) on Fig. 5). The band with the wavenumber 1478 $cm^{-1}$ can be assigned both to the $A_g$ and $B_{1u}$ irreducible representations. Here we take into account, that we used another coordinate system associated with the molecule compared with [22]. Therefore the irreducible representations, which describe the vibrational modes in [22] should be replaced in the following manner ($B_{1u} \rightarrow B_{3u}, B_{2u} \rightarrow B_{1u}, B_{3u} \rightarrow B_{2u}$). The above uncertainty arises from the close values of the wavenumbers, which are obtained and used in [22] for the assignment. The sufficiently strong intensity of this band can be caused by the strong ($Q_{main} - Q_{main}$) and ($d_i - d_i$), $i = (x, y, z)$ contributions in case of belonging to the $A_g$ irreducible representation, or by the ($Q_{main} - d_z$) contributions of vertically or arbitrary oriented molecules in case of belonging to the $B_{1u}$ irreducible representation that causes its strong enhancement also. The line 748 $cm^{-1}$ of the $B_{2u}$ type can be caused by the ($Q_{main} - d_y$) contributions of the horizontally or arbitrary oriented molecules. The line 1354 $cm^{-1}$ of the $B_{3u}$ type can be caused by ($Q_{main} - d_x$) contributions of arbitrary oriented molecules. The line 1522 $cm^{-1}$ of the $B_{2g}$ type can be caused by ($d_x - d_z$) contributions of arbitrary oriented molecules. Analogously we can identify the lines of phenazine at -0.2 V and -0.4 V applied potentials (curve (b) and (c) on Fig. 5), however the number of the lines on the curve (c) is significantly less than at another potentials. As it is mentioned above the most enhancement arises in the first layer of adsorbed molecules. Apparently the effect of the potential is associated with the fact that it prevents significantly penetration of the molecules in the first layer and their binding with the surface, where the strongest enhancement occurs. One should note, that there are some lines, which are shifted significantly with respect to the closest lines referred to some irreducible representations. They are the lines with 632 and 539 $cm^{-1}$. We have some difficulties in assignment of these lines. Apparently they may refer to the protonated phenazine. However we have explained appearance and enhancement of another lines which belong to phenazine, that points out on the validity of our theory.

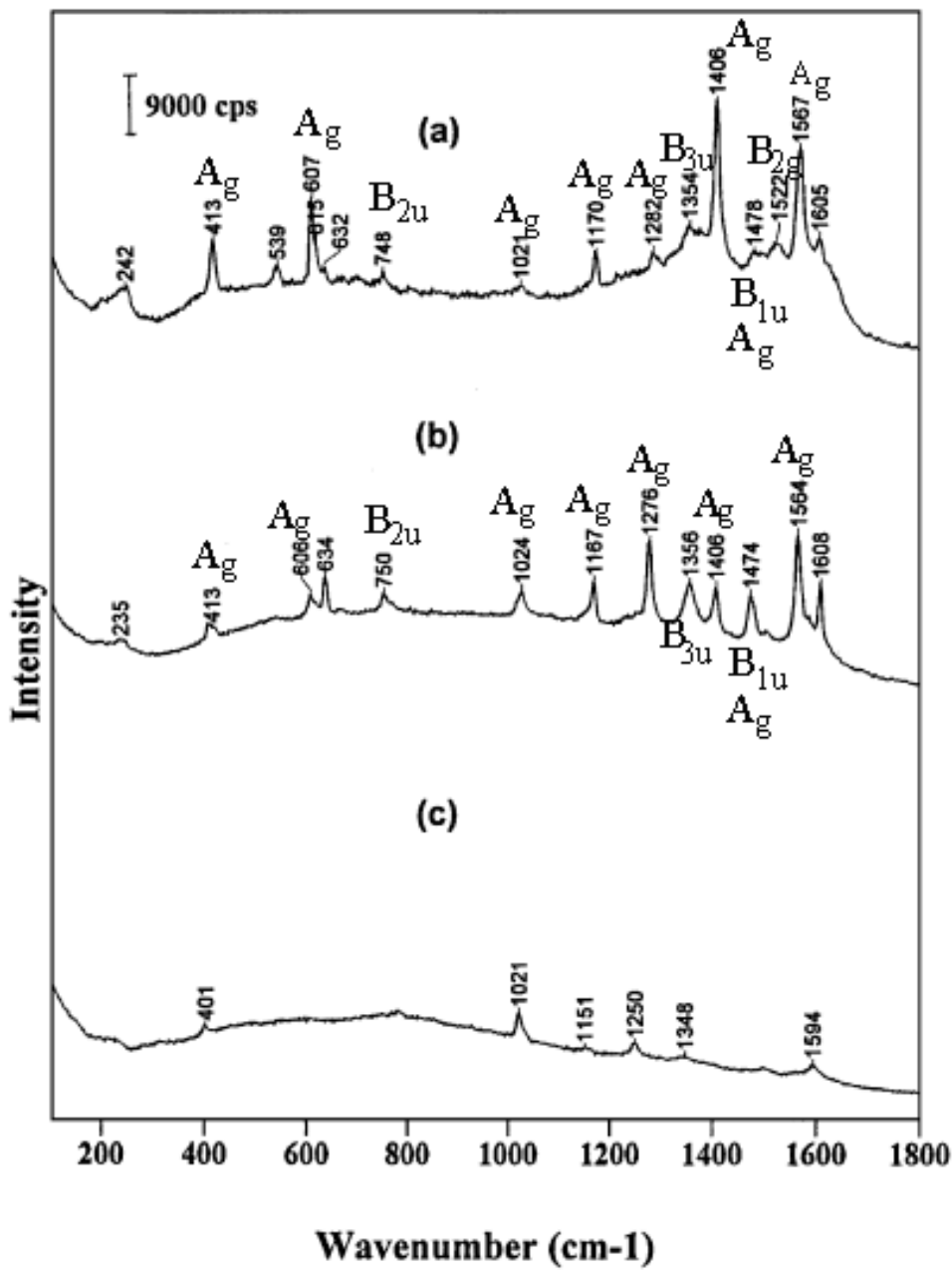

Figure 5 The SER spectrum of phenazine at **(a)** 0 V, **(b)** -0.2 V and **(c)** -0.4 V applied potentials.

Analogously to SERS, the SEHR cross-section has the form [24]

$$d\sigma_{SEHRS\,s,surf}\begin{pmatrix}St\\AnSt\end{pmatrix} = \frac{\omega_{inc}\omega_{scat}^3}{64\pi^2\hbar^4\varepsilon_0^2 c^4}\left|\overline{E}_{inc}\right|_{vol}^2 \frac{\left|\overline{E}_{inc}\right|_{surf}^2 \left|\overline{E}_{inc}\right|_{surf}^2 \left|\overline{E}_{scat}\right|_{surf}^2}{\left|\overline{E}_{inc}\right|_{vol}^2 \left|\overline{E}_{inc}\right|_{vol}^2 \left|\overline{E}_{scat}\right|_{vol}^2} \times$$

$$\begin{pmatrix}\frac{V_s+1}{2}\\\frac{V_s}{2}\end{pmatrix}\left|\sum_{f_1,f_2,f_3} S_{s,f_i-f_j-f_k}\right|^2 dO, \quad (13)$$

Here $S_{s,f_i-f_j-f_k}$ are the scattering contributions for SEHRS, which arise due to the scattering via $f_i$, $f_j$ and $f_k$ dipole and quadrupole moments analogously as in SERS. The scattering contributions obey selection rules [24]

$$\Gamma_s \in \Gamma_{f_i}\Gamma_{f_j}\Gamma_{f_k} \ . \tag{14}$$

The detailed derivation of (13, 14) and the form for the contributions one can find in [24]. Further we shall designate the scattering contributions for SEHRS simply as $(f_i - f_j - f_k)$. In accordance with our estimations of the role of the dipole and quadrupole moments in SERS and SEHRS, the above contributions can be classified after enhancement degree. The sequence below reflects the decrease in the degree of the enhancement for the SEHR scattering contributions.

1. $(Q_{main} - Q_{main} - Q_{main})$ contributions, caused by three main quadrupole moments and their permutations which are most enhanced.

2. $(Q_{main} - Q_{main} - d_{main})$ contributions, caused by two main quadrupole and one main dipole moments and their permutations, which are strongly enhanced too, but with a lesser extent, than the previous ones.

3. $(Q_{main} - d_{main} - d_{main})$ contributions, caused by one main quadrupole moment and by two main dipole moments and their permutations, which can be strongly enhanced, but with a lesser extent than the two previous ones.

4. $(d_{main} - d_{main} - d_{main})$ contributions, caused by three main dipole moments and their permutations, which can be strongly enhanced, but with a lesser extent that the three previous ones.

Other contributions, which contain at least one minor dipole moment apparently are significantly smaller than the above previous ones, while the contributions, which contain minor quadrupole and more minor dipole and quadrupole moments apparently are very small and can not be seen at all. Precise conclusion, concerning the values of these contributions is very difficult.

Since the main quadrupole moments transform after the unit irreducible representation, $\Gamma_s$ for the $(Q_{main} - Q_{main} - Q_{main})$ types of the scattering is unitary also in accordance with the selection rules (14), that points out on the strong enhancement of the bands, caused by totally symmetric vibrations, transforming after the unit irreducible representation. These bands are forbidden in usual HRS for molecules with $C_{nh}$ D and higher symmetry groups. Thus the main feature of the SEHR spectra of symmetrical molecules with these groups is appearance of these strong forbidden bands.

Analysis of experimental SEHR spectrum of pyrazine obtained in [17, 18] also confirms the dipole-quadrupole theory because it allows also to explain appearance of the strong forbidden bands with $A_g$ symmetry. The SEHR spectra, obtained in [17, 18] slightly differ one from another because of some difference in experimental conditions. However the bands symmetry is well known that is sufficient for analysis of the SEHR spectra. Below we shall analyze the SEHR spectrum obtained in [17] (Fig. 4, lower curve (b)). In accordance with the selection rules (11) the lines caused by the vibrations with the following symmetry are observed in the SEHR spectrum:

1. $A_g$ - (636, 1021, 1228, 1593 and 1624 $cm^{-1}$) caused mainly by $(Q_{main} - Q_{main} - Q_{main})$ and $(Q_{main} - d_i - d_i)$, $i = (x, y, z)$ scattering contributions of vertically, horizontally and arbitrary oriented molecules. These lines are forbidden in usual HRS and their appearance strongly proves our point of view.

2. $B_{1u}$ - 1044 $cm^{-1}$ caused mainly by the ($Q_{main} - Q_{main} - d_z$) and ($d_i - d_i - d_z$), $i = (x, y, z)$ scattering contributions of vertically and arbitrary oriented pyrazine. Apparently the contributions ($Q_{main} - Q_{main} - d_z$) and ($d_y - d_y - d_z$) of horizontally adsorbed pyrazine are small since they contain the $d_z$ moment, which is associated with the non enhanced $E_{y'}$ tangential component of the electric field for this orientation. Thus the enhancement of the band with $B_{1u}$ symmetry is caused mainly by the above contributions.

3. $B_{2u}$ – 436 and 796 $cm^{-1}$ caused mainly by the ($Q_{main} - Q_{main} - d_y$) ($d_y - d_y - d_y$) and ($d_i - d_i - d_y$) $i = (x, y, z)$ scattering contributions of horizontally and arbitrary oriented molecules. The vertically adsorbed pyrazine apparently does not determine the intensity of the bands with $B_{2u}$ symmetry, since the corresponding contributions include the non enhanced tangential $E_{y'}$ component of the electric field. Thus the enhancement of the bands with $B_{2u}$ symmetry is caused mainly by the above contributions of horizontally and arbitrary oriented molecules.

4. $B_{3u}$ - 1162 $cm^{-1}$ caused mainly by the ($Q_{main} - Q_{main} - d_x$) and ($d_i - d_i - d_x$), $i = (x, y, z)$ scattering contributions of arbitrary oriented pyrazine.

5. The lines of the $B_{2g}$ and $B_{3g}$ symmetry (1516 and 742 $cm^{-1}$ respectively) may be caused mainly by ($Q_{main} - d_z - d_x$) and ($Q_{main} - d_z - d_y$) of arbitrary oriented molecules. Apparently the vertically and horizontally oriented molecules do not determine the intensities of the bands with $B_{2g}$ and $B_{3g}$ symmetry because of the presence of the non enhanced tangential components of the electric field in corresponding contributions.

The absence of the lines of $A_u$ and $B_{1g}$ symmetry caused by $(d_z - d_x - d_y)$. $(Q_{main} - d_x - d_y)$ and similar scattering contributions may be caused by small enhancement of these lines in the spectrum due to the large number of the components of the electric field, which are not enhanced significantly, that determine not significant value of the enhancement of the above contributions. Thus appearance of all the lines in the SEHR spectrum of pyrazine can be successfully explained by our theory. Now let us explain the main features of the phenazine SEHR spectrum (Fig. 6).

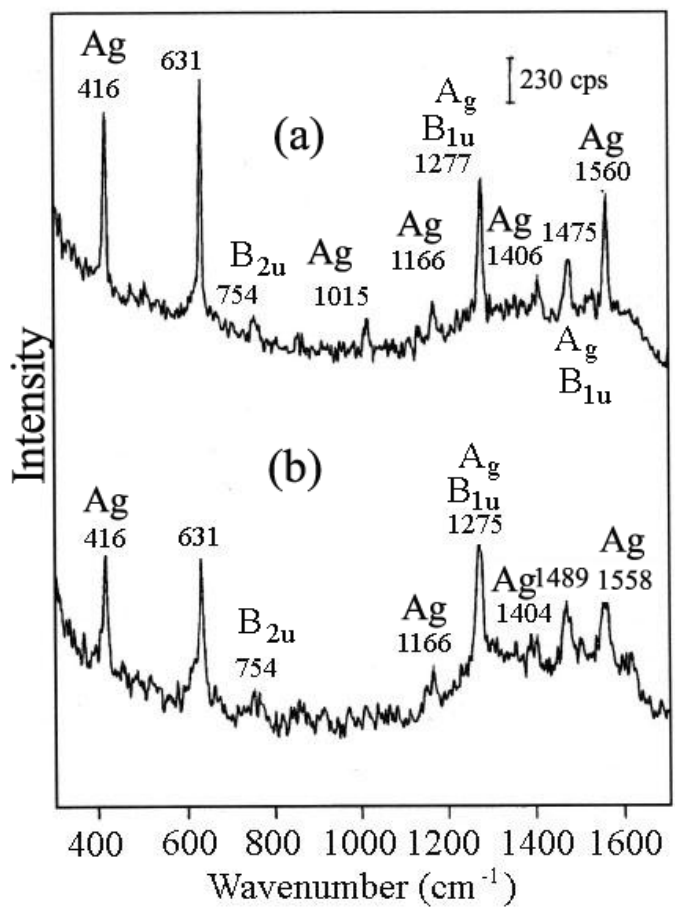

Figure 6  The SEHR spectra of phenazine at **(a)** 0 V and **(b)** − 0.2 V applied potentials.

The main feature of the phenazine SEHR spectra both for 0 V and - 0.2 V applied potentials is appearance of sufficiently strong bands with $A_g$ symmetry which are forbidden in usual HRS in accordance with our theoretical result. They are the bands with 416, 1015, 1166, 1406 and 1560 $cm^{-1}$. Here we write out the wavenumbers, which refer to the 0 V applied potential. The enhancement of these bands is caused by the ($Q_{main} - Q_{main} - Q_{main}$) types of the scatterings and by ($Q_{main} - d_i - d_i$) $i = (x, y, z)$ scattering types of all types of orientations of the molecules. The bands at 1277 $cm^{-1}$ and 1475 $cm^{-1}$ may refer both to the $A_g$ and $B_{1u}$ symmetry types, because of uncertainty of rigorous assignment of these bands associated with close values of the wavenumbers used for the assignment [22]. However, in both cases they can be explained by the ($Q_{main} - Q_{main} - Q_{main}$) and ($Q_{main} - d_i - d_i$) $i = x, y, z$ scattering types for the $A_g$ type of symmetry and by ($Q_{main} - Q_{main} - d_z$) and ($d_i - d_i - d_z$) $i = x, y, z$ scattering types of arbitrary oriented molecules for the $B_{1u}$ type of symmetry. It should be noted that both types of the lines should be strongly enhanced and manifest in the SEHR spectrum of this molecule. The band at 754 $cm^{-1}$ of $B_{2u}$ symmetry type can be explained by the ($Q_{main} - Q_{main} - d_y$) and ($d_i - d_i - d_y$) $i = x, y, z$ scattering types of arbitrary oriented molecules. Thus existence of all the bands of phenazine observed in [21] can be explained by the dipole-quadrupole theory.

One should note that investigation of practically all symmetrical molecules with $C_{nh}$, $D$ and higher symmetry groups reveals existence of strong forbidden lines. Especially they are the lines caused by vibrations with $B_{1u}, B_{2u}, B_{3u}$ irreducible representations in case of SERS and by the vibrations with the unit irreducible representation $A_g$ in case of SEHRS in pyrazine and phenazine, or in the molecules with the $D_{2h}$ symmetry group. The last fact is very important, since it points out the validity of our theory.

# APPENDIX

# THE EXPRESSIONS FOR THE SER CROSS-SECTION AND SELECTION RULES FOR CONTRIBUTIONS

Here we shall obtain an expression for the SER cross –section of symmetrical molecules with the dipole and quadrupole interactions taken into account, which is well applicable to the molecules with $D_{2h}$ and another point symmetry groups with non degenerate electronic and vibrational states. It should be noted that it can be calculated using wave-functions of the time dependent Schrödinger equation.

$$-i\hbar \frac{\partial \Psi}{\partial t} = \left[ \hat{H}_{mol} + \hat{H}_{e-r}^{inc} + \hat{H}_{e-r}^{scat} \right] \Psi \qquad (A-1)$$

where

$$\hat{H}_{mol} = \hat{H}_e + \hat{H}_n + \hat{H}_{e-n} \qquad (A-2)$$

and

$$\hat{H}_e = -\hbar^2/2m \sum_i \Delta_{\bar{r}_i} + \frac{1}{2} \sum_{\substack{i,k \\ i \neq k}} \frac{e^2}{r_{ik}} - \sum_{iJ} \frac{e^2 Z_J^*}{\left|\bar{R}_{iJ}^0\right|} \qquad (A-3)$$

$$\hat{H}_n = -\frac{\hbar^2}{2} \sum_J \frac{1}{M_J} \Delta_{\bar{R}_J} + \frac{1}{2} \sum_{\substack{J,K \\ J \neq K}} \frac{e^2 Z_J^* Z_K^*}{\left|\bar{R}_{JK}\right|} \qquad (A-4)$$

$$\hat{H}_{e-n} = -\sum_{iJ} \frac{e^2 Z_J^*}{\left|\bar{R}_{iJ}\right|} + \sum_{iJ} \frac{e^2 Z_J^*}{\left|\bar{R}_{iJ}^0\right|} \qquad (A-5)$$

Here $\hat{H}_{mol}, \hat{H}_e, \hat{H}_n, \hat{H}_{e-n}$ are the Hamiltonians of the molecule, electrons (in the field of motionless nuclei), nuclei and electron-nuclei interactions. $\bar{r}_i$ is the radius vector of $i$ electron, $r_{ik}$ is the distance between $i$ and $k$ electrons, $\bar{R}_{iJ}^o$ is the radius vector between the motionless $J$ nucleus and $i$ electron, $\overline{R}_{iJ}$ is the radius vector between the $J$ nucleus and $i$ electron, $\overline{R}_J$ is the radius vector of the $J$ nucleus, $\overline{R}_{JK}$ is the radius vector between $J$ and $K$ nuclei, $M_J$ is the mass of the $J$ nucleus, $Z_J^*$ its atomic number. All other designations in ((A-1)-(A-5)) are conventional. The Hamiltonians of interaction of electrons with the incident and scattering fields $\widehat{H}_{e-r}^{inc}$ and $\widehat{H}_{e-r}^{scat}$ in (A-1) are the expression (2) in fact, where we substitute the electric field and the frequency $\omega$ by the values, which refer to the incident and scattered fields respectively

$$\widehat{H}_{e-r}^{inc} = \left|\overline{E}_{inc}\right| \frac{(\overline{e}^*\overline{f}^*)_{inc} e^{i\omega_{inc}t} + (\overline{e}\overline{f})_{inc} e^{-i\omega_{inc}t}}{2} \tag{A-6}$$

$$\widehat{H}_{e-r}^{scat} = \left|\overline{E}_{scat}\right| \frac{(\overline{e}^*\overline{f}^*)_{scat} e^{i\omega_{scat}t} + (\overline{e}\overline{f})_{scat} e^{-i\omega_{scat}t}}{2} \tag{A-7}$$

(It should be noted, that here and further we use for designation of components of vectors and tensors the indices $i, j, k$ and $\alpha, \beta, \gamma$, which mean the values $x, y, z$. In addition the values $x_\alpha, x_\beta, x_\gamma$ designate the coordinates $x, y, z$). In the first stage it is necessary to obtain the molecular wave functions of the unperturbed Hamiltonian, which satisfy the equation

$$\widehat{H}_{mol}\Psi = E\Psi \tag{A-8}$$

This procedure is well described in [1, 2, 25] and we can write the following expressions for the full wave function of the ground state $\Psi_{n\overline{V}}$ and its vibrational part $\alpha_{\overline{V}}$. Taking into account the time dependence, we have

$$\Psi_{n\bar{V}} = \left[ \Psi_n^{(0)} + \sum_{\substack{l \\ l \neq n}} \frac{\sum_s R_{nls} \sqrt{\frac{\omega_s}{\hbar}} \xi_s \Psi_l^{(0)}}{(E_n^{(0)} - E_l^{(0)})} \right] \alpha_{\bar{V}} \exp-(iE_{n\bar{V}}t)/\hbar \qquad (A-9)$$

$$\alpha_{\bar{V}} = \prod_s N_s H_{V_s}\left(\sqrt{\frac{\omega_s}{\hbar}}\xi_s\right) \exp\left(-\frac{\omega_s \xi_s^2}{2\hbar}\right) \qquad (A-10)$$

Here the full wave function is characterized by the electron wave number $n$ and by the set of the vibration wave numbers $\bar{V} = (V_1, V_2 ... V_s ...)$, which consists of the wave numbers of separate vibrations $V_s$. $\Psi_n^{(0)}$ and $\Psi_l^{(0)}$ are the solutions for the ground and excited states of the equation

$$\hat{H}_e \Psi^{(0)} = E^{(0)} \Psi^{(0)} \qquad (A-11)$$

$E_n^{(0)}$ and $E_l^{(0)}$ are their energies. $\xi_s$ is the normal coordinate of the $s$ vibration mode.

$$R_{nls} = \sum_{J\alpha} \sqrt{\frac{\hbar}{\omega_s}} \bar{X}_{Js\alpha} \frac{\partial \langle l|\hat{H}_{e-n}|n \rangle}{\partial X_{J\alpha}} \qquad (A-12)$$

is the coefficient, which characterizes excitation of the $\Psi_l^{(0)}$ wave function by the $s$ vibrational mode from the ground state $n$ to the excited states $l$, $\omega_s$ is a frequency of the $s$ vibrational mode, $\bar{X}_{Js\alpha}$ is an $\alpha$ component of the displacement vector of the $J$ nucleus in the $s$ vibrational mode, $X_{J\alpha}$ is an $\alpha$ component of the coordinate of the $J$ nucleus. Here and further we mean, that the derivatives of the matrix elements of $\hat{H}_{e-n}$ are taken for $X_{J\alpha}^0$ values which correspond to the equilibrium positions of nuclei. $N_s$ and $H_{V_s}$ in (34a) are normalization constants and the Hermitian polynomials for separate vibrations.

$$E_{n\bar{V}} = E_n^{(0)} + \sum_s \hbar\omega_s (V_s + 1/2) \tag{A-13}$$

is the expression for the full energy of the state. The specific form for the time dependent perturbed wave function, which takes into account the interaction of light with the molecule (A-6), (A-7) can be obtained from the general expression for this function in accordance with [26].

$$\Psi_{n\bar{V}}^{(1)}(t) = \sum_{k\bar{V}_1} a_{(k\bar{V}_1),(n\bar{V})}(t) \Psi_{k\bar{V}_1} \tag{A-14}$$

Here $\Psi_{n\bar{V}}^{(1)}(t)$ is the time dependent perturbated wave function, $\Psi_{k\bar{V}_1}$ are the eigenfunctions of the unperturbed Hamiltonian. $\bar{V}_1$ are sets of vibrational wave numbers of excited states. The cross-section of Raman scattering is expressed via the second terms of the expansions of the perturbation coefficients $a_{(n,\bar{V}\pm 1),(n,\bar{V})}^{(2)}(t)$, where the index $\bar{V} \pm 1$ designates the change of one of vibrational numbers $V_s$ on one and refer to the Stokes and AntiStokes scattering respectively. Since the coefficients $a_{(n,\bar{V}\pm 1),(n,\bar{V})}^{(2)}(t)$ are very small, the full electron wave function is determined mainly by the wave function of the ground state. We neglect by the difference between the vibrational functions and vibrational frequencies of the ground and virtual electron states that strongly simplified our calculations. The wave functions of the virtual states can be easily determined by generalization of the procedure, used to obtain $\Psi_{n\bar{V}}$ in [1, 2, 25] by changing of indices. If we use the above mentioned concept concerning the vibrational states and frequencies the virtual wave functions $\Psi_{m,\bar{V}}$ have the form

$$\Psi_{m\overline{V}} = \left[ \Psi_m^{(0)} + \sum_{\substack{k \\ k \neq m}} \frac{\sum_s R_{mks} \sqrt{\frac{\omega_s}{\hbar}} \xi_s \Psi_k^{(0)}}{(E_m^{(0)} - E_k^{(0)})} \right] \alpha_{\overline{V}} \exp-(iE_{m\overline{V}}t)/\hbar \qquad (A\text{-}15)$$

Here

$$R_{mks} = \sum_{J\alpha} \sqrt{\frac{\hbar}{\omega_s}} \overline{X}_{Js\alpha} \frac{\partial \langle k | \hat{H}_{e-n} | m \rangle}{\partial X_{J\alpha}} \qquad (A\text{-}16)$$

similarly to $R_{nls}$ (A-12). The cross-section of Raman scattering is expressed in terms of the expression

$$a^{(2)}_{(n,\overline{V}\pm 1),(n,\overline{V})}(t) = \left(-\frac{i}{\hbar}\right)^2 \sum_{\substack{m \\ m \neq n}} \left[ \int_0^t \langle n, \overline{V}\pm 1 | \hat{H}^{inc}_{e-r} + \hat{H}^{scat}_{e-r} | m, \overline{V}\pm 1 \rangle dt_1 \times \right.$$

$$\times \int_0^{t_1} \langle m, \overline{V}\pm 1 | \hat{H}^{inc}_{e-r} + \hat{H}^{scat}_{e-r} | n, \overline{V} \rangle dt_2 + \qquad (A\text{-}17)$$

$$\left. + \int_0^t \langle n, \overline{V}\pm 1 | \hat{H}^{inc}_{e-r} + \hat{H}^{scat}_{e-r} | m, \overline{V} \rangle dt_1 \int_0^{t_1} \langle m, \overline{V} | \hat{H}^{inc}_{e-r} + \hat{H}^{scat}_{e-r} | n, \overline{V} \rangle dt_2 \right]$$

Taking into account the expressions for the Hamiltonians of interaction of light with electrons (A-6),(A-7) and expressions (A-9),(A-15) for the wave functions of the ground and excited states one can obtain the following expression for the Raman cross-section in a free space

$$d\sigma_{vol} = \frac{\omega_{inc} \omega_{scat}^3}{16\hbar^2 \varepsilon_0^2 \pi^2 c^4} \times \begin{pmatrix} (V_s+1)/2 \\ V_s/2 \end{pmatrix} \left| A_{V_s} \begin{pmatrix} St \\ anSt \end{pmatrix} \right|^2_{vol} dO \qquad (A\text{-}18)$$

where

$$A_{V_s}\begin{pmatrix} St \\ anSt \end{pmatrix} = \sum_{\substack{m,l \\ m,l \neq n}} \frac{R_{nls} \langle n | (\overline{e}^* \overline{f}_e^*)_{scat} | m \rangle \langle m | (\overline{e} \overline{f}_e)_{inc} | l \rangle}{(E_n^{(0)} - E_l^{(0)})(\omega_{mn} \pm \omega_s - \omega_{inc})} +$$

$$\sum_{\substack{m,l \\ m,l \neq n}} \frac{R_{nls}^* \langle l|(\bar{e}^*\bar{f}_e^*)_{scat}|m\rangle\langle m|(\bar{e}\bar{f}_e)_{inc}|n\rangle}{(E_n^{(0)} - E_l^{(0)})(\omega_{mn} - \omega_{inc})} +$$

$$\sum_{\substack{m,l \\ m,l \neq n}} \frac{R_{nls} \langle n|(\bar{e}\bar{f}_e)_{inc}|m\rangle\langle m|(\bar{e}^*\bar{f}_e^*)_{scat}|l\rangle}{(E_n^{(0)} - E_l^{(0)})(\omega_{mn} \pm \omega_s + \omega_{scat})} +$$

$$\sum_{\substack{m,l \\ m,l \neq n}} \frac{R_{nls}^* \langle l|(\bar{e}\bar{f}_e)_{inc}|m\rangle\langle m|(\bar{e}^*\bar{f}_e^*)_{scat}|n\rangle}{(E_n^{(0)} - E_l^{(0)})(\omega_{mn} + \omega_{scat})} +$$

$$\sum_{\substack{m,k \\ m \neq n,k}} \frac{\langle n|(\bar{e}^*\bar{f}_e^*)_{scat}|m\rangle R_{mks}^* \langle k|(\bar{e}\bar{f}_e)_{inc}|n\rangle}{(E_m^{(0)} - E_k^{(0)})(\omega_{mn} \pm \omega_s - \omega_{inc})} +$$

$$\sum_{\substack{m,k \\ m \neq n,k}} \frac{\langle n|(\bar{e}^*\bar{f}_e^*)_{scat}|k\rangle R_{mks} \langle m|(\bar{e}\bar{f}_e)_{inc}|n\rangle}{(E_m^{(0)} - E_k^{(0)})(\omega_{mn} - \omega_{inc})} +$$

$$\sum_{\substack{m,k \\ m \neq n,k}} \frac{\langle n|(\bar{e}\bar{f}_e)_{inc}|m\rangle R_{mks}^* \langle k|(\bar{e}^*\bar{f}_e^*)_{scat}|n\rangle}{(E_m^{(0)} - E_k^{(0)})(\omega_{mn} \pm \omega_s + \omega_{scat})} +$$

$$\sum_{\substack{m,k \\ m \neq n,k}} \frac{\langle n|(\bar{e}\bar{f}_e)_{inc}|k\rangle R_{mks} \langle m|(\bar{e}^*\bar{f}_e^*)_{scat}|n\rangle}{(E_m^{(0)} - E_k^{(0)})(\omega_{mn} + \omega_{scat})} + \quad \text{(A-19)}$$

Here

$$\omega_{mn} = \frac{E_m^{(0)} - E_n^{(0)}}{\hbar} \quad \text{(A-20)}$$

The SER cross-section can be obtained from expression (A-18) with taking into account of the fact that the incident field affecting the molecule is the surface field. The same refers to the scattered field. Therefore one should multiply (A-18) by the expression

$$\frac{|\overline{E}_{inc}|^2_{surf}}{|\overline{E}_{inc}|^2_{vol}} \frac{|\overline{E}_{scat}|^2_{surf}}{|\overline{E}_{scat}|^2_{vol}} \tag{A-21}$$

and take the surface fields as the incident and scattered fields under the sign of modulus in (A-18). Finally the expression for the SER cross-section has the form

$$d\sigma_{surf} = \frac{\omega_{inc}\omega_{scat}^3}{16\hbar^2 \varepsilon_0^2 \pi^2 c^4} \frac{|\overline{E}_{inc}|^2_{surf}}{|\overline{E}_{inc}|^2_{vol}} \frac{|\overline{E}_{scat}|^2_{surf}}{|\overline{E}_{scat}|^2_{vol}} \times$$
$$\times \binom{(V_s+1)/2}{V_s/2} \left| A_{V_s}\binom{St}{anSt} \right|^2_{surf} dO \tag{A-22}$$

Here the signs *surf* and *vol* designate the values of the cross-sections and electromagnetic fields at the surface and in the volume. It should be noted, that $(\overline{E}_{inc})_{surf}$ is the surface field generated by the field $(\overline{E}_{inc})_{vol}$ incident on the surface, while $(\overline{E}_{scat})_{surf}$ is the surface field generated by the field $(\overline{E}_{scat})_{vol}$ incident on the surface from the direction for which the direction of the wave reflected from the surface coincides with the direction of the scattering. Let us introduce the scattering tensor $C_{V_s,}[f_i, f_j]$, which has a form

$$C_{V_s}[f_i, f_j] = \sum_{\substack{m,l \\ m,l \neq n}} \frac{R_{nls}\langle n|f_i|m\rangle\langle m|f_j|l\rangle}{(E_n^{(0)} - E_l^{(0)})(\omega_{mn} \pm \omega_s - \omega_{inc})} +$$

$$\sum_{\substack{m,l \\ m,l \neq n}} \frac{R_{nls}^*\langle l|f_i|m\rangle\langle m|f_j|n\rangle}{(E_n^{(0)} - E_l^{(0)})(\omega_{mn} - \omega_{inc})} +$$

$$\sum_{\substack{m,l \\ m,l \neq n}} \frac{R_{nls}\langle n|f_j|m\rangle\langle m|f_i|l\rangle}{(E_n^{(0)} - E_l^{(0)})(\omega_{mn} \pm \omega_s + \omega_{scat})} +$$

$$\sum_{\substack{m,l \\ m,l \neq n}} \frac{R_{nls}^{*}\langle l|f_j|m\rangle\langle m|f_i|n\rangle}{(E_n^{(0)} - E_l^{(0)})(\omega_{mn} + \omega_{scat})} +$$

$$\sum_{\substack{m,k \\ m \neq n,k}} \frac{\langle n|f_i|m\rangle R_{mks}^{*}\langle k|f_j|n\rangle}{(E_m^{(0)} - E_k^{(0)})(\omega_{mn} \pm \omega_s - \omega_{inc})} +$$

$$\sum_{\substack{m,k \\ m \neq n,k}} \frac{\langle n|f_i|k\rangle R_{mks}\langle m|f_j|n\rangle}{(E_m^{(0)} - E_k^{(0)})(\omega_{mn} - \omega_{inc})} +$$

$$\sum_{\substack{m,k \\ m \neq n,k}} \frac{\langle n|f_j|m\rangle R_{mks}^{*}\langle k|f_i|n\rangle}{(E_m^{(0)} - E_k^{(0)})(\omega_{mn} \pm \omega_s + \omega_{scat})} +$$

$$\sum_{\substack{m,k \\ m \neq n,k}} \frac{\langle n|f_j|k\rangle R_{mks}\langle m|f_i|n\rangle}{(E_m^{(0)} - E_k^{(0)})(\omega_{mn} + \omega_{scat})} \tag{A-23}$$

where under $f_i$ and $f_j$ we mean the dipole and quadrupole moments. Using the obvious property

$$C_{V_{(s,p)}}[f_i,(a_1 f_j + a_2 f_k)] = a_1 C_{V_{(s,p)}}[f_i, f_j] + a_2 C_{V_{(s,p)}}[f_i, f_k]$$

$$C_{V_{(s,p)}}[(a_1 f_i + a_2 f_j), f_k] = a_1 C_{V_{(s,p)}}[f_i, f_k] + a_2 C_{V_{(s,p)}}[f_j, f_k] \tag{A-24}$$

the expression for $A_{V_s}$ can be presented as a sum of several contributions

$$A_{V_s} = \sum_{f_i, f_k} S_{s, f_i - f_j}, \tag{A-25}$$

The expression for $\sum_{f_i, f_j} S_{s, f_i - f_j}$ consists from various contributions of $(d-d)$ scattering type

$$S_{s,d-d} = \sum_{i,k} C_{V_s}[d_i, d_k] (e_{scat,i}^{*} e_{inc,k})_{surf} \tag{A-26}$$

of $(d-Q)$ scattering type

$$S_{s,d-Q} = \sum_{i,\gamma\delta} C_{V_s}[d_i, Q_{\gamma\delta}] \left( e^*_{scat,i} \frac{1}{2|\bar{E}_{inc}|} \frac{\partial E^{inc}_\gamma}{\partial x_\delta} \right)_{surf} \qquad (A-27)$$

of $(Q-d)$ scattering type

$$S_{s,Q-d} = \sum_{\alpha\beta,k} C_{V_{(s,p)}}[Q_{\alpha\beta}, d_k] \left( \frac{1}{2|\bar{E}_{scat}|} \frac{\partial E^{scat^*}_\alpha}{\partial x_\beta} e_{inc,k} \right) \qquad (A-28)$$

and of the $(Q-Q)$ scattering type

$$S_{s,Q-Q} = \sum_{\alpha\beta,\gamma\delta} C_{V_s}[Q_{\alpha\beta}, Q_{\gamma\delta}] \left( \frac{1}{2|\bar{E}_{scat}|} \frac{\partial E^{scat^*}_\alpha}{\partial x_\beta} \frac{1}{2|\bar{E}_{inc}|} \frac{\partial E^{inc}_\gamma}{\partial x_\delta} \right)_{surf} \qquad (A-29)$$

The scattering contributions $S_{s,f_i-f_j}$ obey selection rules which can be obtained from the expression for the scattering tensor (A-23), when the condition

$$C_{V_{(s,p)}}[f_i, f_j] \neq 0 \qquad (A-30)$$

is fulfilled. Let us first consider the first term of (A-23) under condition, that

$$R_{nls} \langle n|f_i|m \rangle \langle m|f_j|l \rangle \neq 0 \; . \qquad (A-31)$$

This expression is valid, when

$$R_{nls} \neq 0 \; , \qquad (A-32)$$

$$\langle n|f_i|m \rangle \neq 0 \; , \qquad (A-33)$$

$$\langle m|f_j|l \rangle \neq 0 \qquad (A-34)$$

satisfy simultaneously. It can be demonstrated [1], that the expression (A-32) is valid, when

$$\Gamma_l \in \Gamma_s \Gamma_n \; , \qquad (A-35)$$

where the sign $\Gamma$ designates the irreducible representations of the $s$ vibration and of the $n$ and $l$ electronic states. Thus this expression determines the irreducible representation of the $l$ excited state, caused by the $s$ vibration. In accordance with general principles of the group theory [27] the expressions (A-33) and (A-34) are valid under conditions, that

$$\Gamma_m \in \Gamma_{f_j}\Gamma_l \ , \tag{A-36}$$

$$\Gamma_n \in \Gamma_{f_i}\Gamma_m . \tag{A-37}$$

Using the expressions ((A-35)-(A-37)) one can obtain the final expression

$$\Gamma_s \in \Gamma_{f_i}\Gamma_{f_j} , \tag{A-38}$$

Analysis of other lines in (A-23) results in the same expression (A-38). Thus (A-38) is valid for the whole tensor and determines the selection rules for the contributions with $f_i$ and $f_j$ dipole and quadrupole moments.